\documentclass[12pt]{article}

\title{Quantum Coherence in Photo-Ionization with Tailored XUV Pulses}
\date{\today}

\usepackage{authblk}

\author[1,*]{Stefanos~Carlstr\"{o}m}
\author[1]{Johan~Mauritsson}
\author[2]{Kenneth~J.~Schafer}
\author[1]{Anne~L'Huillier}
\author[1,$\dagger$]{Mathieu~Gisselbrecht}
\affil[1]{Department of Physics, Lund University, Box 118, 221 10 Lund, Sweden}
\affil[2]{Department of Physics and Astronomy, Louisiana State University, Baton Rouge, LA 70803}
\affil[*]{Email: stefanos.carlstrom@gmail.com}
\affil[$\dagger$]{Email: mathieu.gisselbrecht@sljus.lu.se}

\input{defs.sty}
\begin{document}

\maketitle
\vspace{-2ex}

\label{sec:abstract}
\begin{abstract}
  \noindent Ionization with ultrashort pulses in the extreme
  ultraviolet (XUV) regime can be used to prepare an ion in a
  superposition of spin--orbit substates. In this work, we study the
  coherence properties of such a superposition, created by ionizing
  xenon atoms using two phase-locked XUV pulses at different
  frequencies. In general, if the duration of the driving pulse
  exceeds the quantum beat period, dephasing will occur. If however,
  the frequency difference of the two pulses matches the spin--orbit
  splitting, the coherence can be efficiently increased and dephasing
  does not occur.
\end{abstract}
\label{sec:org93bc61a}
\section{Introduction}
\label{sec:introduction}
\begin{figure}[tb]
  \centering
\tikzsetnextfilename{comp-systems}
\begin{tikzpicture}[scale=0.8]
  \begin{scope}
    \draw (0,0) -- ++(2,0) node[anchor=west]{\(\ket{1}\)};
    \draw (0,1.5) -- ++(2,0) node[anchor=west]{\(\ket{2}\)};
    \draw[->,color=accent] (1,0) -- ++(0,1.5);
    \node at (-1,2) {(a)};
  \end{scope}
  \begin{scope}[xshift=5.5cm]
    \draw (0,0) -- ++(2,0) node[anchor=west]{\(\ket{0}\)};
    \draw (0,1.5) -- ++(2,0) node[anchor=west]{\(\ket{1}\)};
    \draw (0,2) -- ++(2,0) node[anchor=west]{\(\ket{2}\)};
    \draw[->,color=accent] (1,0) -- ++(0,1.75);
    \draw[domain=1.25:2.25,smooth,variable=\y,black]  plot ({exp(-(\y-1.75)^2/(2*0.5^2))+0.5},{\y});
    \node at (-1,2) {(b)};
  \end{scope}
  \begin{scope}[xshift=11cm]
    \fill[pattern = north east lines, pattern color=dimmed] (0,2.0) rectangle ++(2,0.2);
    \draw (0,0) -- ++(2,0) node[anchor=west]{\(\ket{0}\)};
    \fill[pattern = north east lines, pattern color=dimmed] (0,1.5) rectangle ++(2,0.2);
    \draw (0,1.5) -- ++(2,0) node[anchor=west]{\(\ket{1}\)};
    \draw (0,2) -- ++(2,0) node[anchor=west]{\(\ket{2}\)};
    \draw[->,color=accent] (1,0) -- (1,0.6*6);
    \draw[domain=-0.5:0.5,smooth,variable=\y,black]  plot ({exp(-(\y)^2/(2*0.5^2))+0.5},{\y+1.75});
    \node at (-1,2) {(c)};
  \end{scope}
  \begin{scope}[xshift=16.5cm]
    \fill[pattern = north east lines, pattern color=dimmed] (0,2.0) rectangle ++(2,0.2);
    \draw (0,0) -- ++(2,0) node[anchor=west]{\(\ket{0}\)};
    \fill[pattern = north east lines, pattern color=dimmed] (0,1.5) rectangle ++(2,0.2);
    \draw (0,1.5) -- ++(2,0) node[anchor=west]{\(\ket{1}\)};
    \draw (0,2) -- ++(2,0) node[anchor=west]{\(\ket{2}\)};
    \foreach \y in {1,2,...,4}{
      \draw[->,color=accent] (1,{0.6*(\y-1)}) -- (1,{0.6*\y});
    }
    \draw[domain=-0.5:0.5,smooth,variable=\y,black]  plot ({exp(-(\y)^2/(2*0.5^2))+0.5},{\y+1.75});
    \node at (-1,2) {(d)};
  \end{scope}
\end{tikzpicture}
  \caption{\label{fig:comp-systems}~Different ways of preparing
    coherent superpositions using light; (a) excitation of a two-level
    system, such as those used for qubits in quantum information; (b)
    coherent excitation from the ground state to two excited bound
    states; (c) single-photon ionization, with the ion left in a
    superposition of substates; (d) strong-field ionization, also
    leaving the ion in a superposition of substates.}
\end{figure}
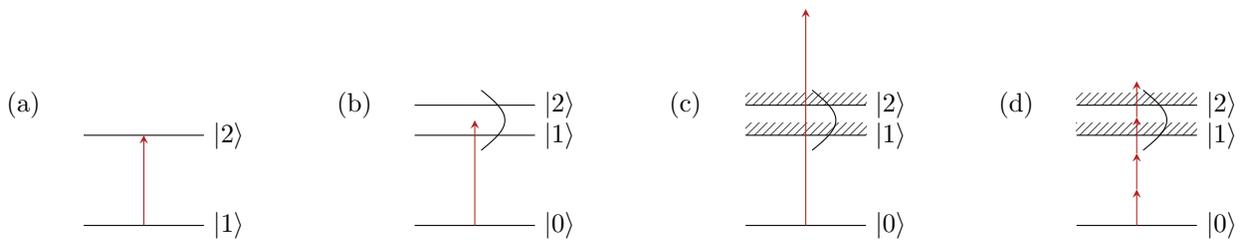
The wave nature of matter is central to the quantum mechanical
description of the microcosmos; therefore coherence --- a measure of the
ability to produce stationary interference patterns --- is an important
property of any quantum system. An example of a coherent system is the
superposition of two pure states,
\(\ket{\psi} = a\ket{1} + b\ket{2}\); such superpositions form the basis
for the field of quantum information, where they are used to represent
qubits. The manipulation of qubits for quantum computing necessarily
requires that the coherence of the system is retained; if not, the
information contained within the qubit is lost. In quantum optics,
superpositions between two states may be created via a transition
between the two states with an appropriately tailored pulse (\eg a
\(\cpi/2\)-pulse) [figure~\ref{fig:comp-systems}~(a)].

\begin{figure}[tb]
  \begin{floatrow}
    \floatbox{table}[][][t]{\caption{\label{tab:quantum-beat-periods}
        Some properties of heavy noble gases.
        \(\Ddiff E_{\textrm{s--o}}\) is the spin--orbit splitting of the
        ionic ground state \(n\conf{p}^5\;\conf{^2P^o}\). The quantum
        beat period, \(T (=2\cpi\Ddiff E_{\textrm{s--o}}^{-1}\)), is
        the intrinsic atomic clock associated to two states separated
        by an energy difference of \(\Ddiff E_{\textrm{s--o}}\).}}{
      \footnotesize
      \begin{tabular}{rrrlr}
        Element & \(Z\) & \(n\)
        & \(\Ddiff E_{\textrm{s--o}}\) [\si{\electronvolt}]
        & \(T\) [\si{\femto\second}]\\
        \hline
        Ne & 10 & 2 & 0.09676024 & 42.8\\
        Ar & 18 & 3 & 0.17749368  & 23.3\\
        Kr & 36 & 4 & 0.665808  & 6.2\\
        Xe & 54 & 5 & 1.306423 & 3.2\\
      \end{tabular}
    }
    \floatbox{figure}[][][t]{\caption{\label{fig:quantum-path-overlap}
        Overlap (shaded area) between two quantum paths (1 and 2)
        separated by \(\Ddiff E\). The spectral bandwidth, \(\Omega\), of
        the ionizing pulse is inversely proportional to the duration
        of the pulse; increasing the pulse duration thus leads to
        decreased overlap between the quantum
        paths. 
        }} {
\tikzsetnextfilename{coherence-uncertainty}
\begin{tikzpicture}
  \fill[domain=-1:1,samples=100,smooth,variable=\x,black!20!white] plot
  (\x,{exp(-(\x + (\x > 0 ? 1 : -1))^2/0.6)});
  \draw[->] (-4,0) -- (4,0) node[anchor=north east] {Energy};
  \draw[dimmed] (0,-0.1) -- (0,1.3);
  \draw[dimmed] (1,-0.1) -- (1,1.1);
  \draw[dimmed] (-1,-0.1) -- (-1,1.1);
  \draw[domain=-3:3,samples=200,smooth,variable=\x,accent] plot (\x,{exp(-(\x+1)^2/0.6)});
  \node[anchor=south] at (-1,1.2) {$1$};
  \draw[domain=-3:3,samples=200,smooth,variable=\x,accent] plot (\x,{exp(-(\x-1)^2/0.6)});
  \node[anchor=south] at (1,1.2) {$2$};
  \draw[decorate,decoration=brace] (1,-0.2) -- ++(-2,0)
  node[anchor=north,midway]{\(\Ddiff E\)};
  \fill[white] (-1.1,0.2) rectangle ++(0.2,0.55);
  \draw[<->,accent] (-1.8,0.3) -- ++ (1.6,0) node[anchor=south,midway]{\(\Omega\)};
\end{tikzpicture}}
  \end{floatrow}
\end{figure}
Superpositions of states can also be achieved by direct excitation
using short light pulses [figure~\ref{fig:comp-systems}~(b)], provided
the bandwidth of the pulse is larger than the energy difference
(\(\omega_{21}\)) between the two states.  This requires a pulse duration
short enough, \(\tau \leq {2\cpi}/{\omega_{21}}\). If the superposition is
successfully created, it may be observed through quantum beats
\parencite{Teets1977,Salour1977,Mauritsson2010PRL,Tzallas2011NP} which
usually last substantially longer than the pulse duration. The
characteristic decay time is termed the \emph{coherence time}. In the
cases depicted in \ref{fig:comp-systems}~(a) and (b), the light
couples the bound states and enables coherent population transfer.

Another way to produce a superposition of states is via short-pulse
ionization, when the ion is left in a coherent superposition of final
states, \eg due to spin--orbit interaction. This can be done using
either high-frequency [figure~\ref{fig:comp-systems}~(c)] or
high-intensity short-pulse [figure~\ref{fig:comp-systems}~(d)]
radiation. As previously, the bandwidth of the ionizing pulse has to
exceed the energy splitting between the ion
states. \textcite{Kurka2009} investigated case \ref{fig:comp-systems}
(c) by photo-ionizing neon using short XUV pulses from a free-electron
laser. A coherent superposition of the ionic fine-structure substates
was prepared and probed by subsequent ionization. Using a strong laser
field [figure~\ref{fig:comp-systems}~(d)], \textcite{Goulielmakis2010}
photo-ionized krypton, leaving the residual Kr\(^+\) ion in a coherent
superposition of the ionic substates. The quantum beat was observed by
probing with a delayed attosecond (\si{\atto\second}) XUV pulse, as
long as the duration of the ionizing pulse was shorter than
the quantum beat period of the Kr\(^+\) ion.
This experimental activity stimulated an important theoretical effort
to investigate the coherence of superpositions of states produced
either directly by photo-excitation [figure~\ref{fig:comp-systems}
(b); \cite{Tzallas2011NP,Klunder2013PRA}], single-photon ionization
[figure \ref{fig:comp-systems}~(c); \cite{Nikolopoulos2013PRL}], or
strong-field ionization [figure~\ref{fig:comp-systems}~(d);
\cite{Pabst2011PRL,Pabst2016}].

As mentioned above, the creation of a coherent superposition in the
cases depicted in figures~\ref{fig:comp-systems}~(b--d), requires
sufficient bandwidth of the exciting/ionizing radiation. As the pulse
duration increases, or the energy separation between electronic states
increases, states become spectrally resolvable. Excitation/ionization
to one state or the other can then be seen as distinguishable quantum
paths taken by the system (see
figure~\ref{fig:quantum-path-overlap}). When the spectral overlap
between these quantum paths decreases, the coherence between states
diminishes. While a spectral representation provides meaningful
physical insight, it does not allow understanding how coherence is
built-up in real time. Therefore, a temporal representation, based on
an atomic clock constituted by the quantum beat period, is very useful
to determine whether decoherence will occur or not during light--matter
interaction. As long as the pulse duration is shorter than a quantum
beat period, the atomic clock will not dephase, since there is still
appreciable overlap between the quantum paths (see
figure~\ref{fig:quantum-path-overlap}). For ionization of noble gases,
the available quantum beat periods span an order of magnitude (see
table~\ref{tab:quantum-beat-periods}). However, ultrashort pulses are
still necessary to manipulate the coherence.

In this article, we present a theoretical study of single-photon
ionization of xenon atoms using XUV pulses, tailored in such a way
that the quantum paths \emph{always} overlap spectrally, providing a
new ionization scheme to control and maximize the coherence between
states, independent of the pulse duration. This is achieved by
employing a bichromatic or multi-colour ionizing field, consisting of
phase-locked harmonics such as those resulting from high-order
harmonic generation (HHG), provided the spectral components fulfill a
certain resonance condition. We investigate the tolerance of this
resonance condition, \ie how strict the requirements on the driving
field are, with respect to the excitation frequencies, pulse duration,
and temporal structure, for maintaining a certain level of
coherence. We find the existence of resonant conditions, which
correspond to a situation where multiple quantum paths lead to the
same photo-electron state.

The paper is organized as follows; in the following section, we
present our numerical calculations based on the fully correlated
time-dependent Schrödinger equation (TDSE) in the case of weak-field
ionization, as well as the theoretical tools that are used to
calculate the evolution of the superposition of states in the presence
of the driving field. We also introduce the density matrix formalism
used to analyze the coherence of the quantum system. Atomic units are
used throughout, unless otherwise stated. Using the numerical
calculations, in section~\ref{sec:bichromatic}, we study ionization
using two harmonic components, and investigate the relation between
their energy separation and the spin--orbit splitting using a temporal
model of the ionization dynamics. From this, we extract a generalized
quantum beat condition. Finally in section~\ref{sec:multi-colour}, we
derive a spectral model for the general case of ionization by
multi-colour fields. This enables us to capture the essential physics
of the observed phenomena. We conclude with a short discussion about
the foundation of the work in relation to quantum mechanics,
suggesting that coherence between states should exist as long as
ionization pathways are indistinguishable by the measurement.

\section{Theoretical framework}
\label{sec:theoretical-framework}
\noindent We are interested in studying the coherence of different
ionic states produced by photo-ionization with tailored XUV
pulses. To this end, we use as a model system noble gas ions, which
have a spin--orbit splitting of the ground state
(\(n\conf{p^5\;^2P}^\conf{o}_{j_i}\),
\(j_i=\sfrac{3}{2},\sfrac{1}{2}\)), in particular xenon (\(n=5\)).

\begin{figure}[tb]
  \begin{floatrow}
    \floatbox{table}[\FBwidth][][t]{\caption{\label{tab:ion-channels}
        Ionization channels accessible via one-photon ionization from
        the valence shell of a noble gas (final \(J=1\)), in the case
        of \(jK\) coupling.}}{
      \begin{tabular}{r|rl}
        \# & Channel configuration\\
        \hline
        1 & \(n\conf{p}^5(\conf{^2P^o_{\sfrac{3}{2}}})k\conf{d}\;^2[\sfrac{1}{2}]_{1}\)\Tstrut\\
        2 & \(n\conf{p}^5(\conf{^2P^o_{\sfrac{3}{2}}})k\conf{s}\;^2[\sfrac{3}{2}]_{1}\)\\
        3 & \(n\conf{p}^5(\conf{^2P^o_{\sfrac{3}{2}}})k\conf{d}\;^2[\sfrac{3}{2}]_{1}\)\\
        \hline
        4 & \(n\conf{p}^5(\conf{^2P^o_{\sfrac{1}{2}}})k\conf{s}\;^2[\sfrac{1}{2}]_{1}\)\Tstrut\\
        5 & \(n\conf{p}^5(\conf{^2P^o_{\sfrac{1}{2}}})k\conf{d}\;^2[\sfrac{3}{2}]_{1}\)\\
        \hline
        6 & \(n\conf{s}n\conf{p}^6(\conf{^2S_{\sfrac{1}{2}}})k\conf{p}\;^2[\sfrac{1}{2}]_{1}\)\Tstrut\\
        7 & \(n\conf{s}n\conf{p}^6(\conf{^2S_{\sfrac{1}{2}}})k\conf{p}\;^2[\sfrac{3}{2}]_{1}\)\\
      \end{tabular}}
    \floatbox{figure}[\Xhsize][][t]{\caption{\label{fig:sketch} Schematic
        energy diagram of a noble gas (heavier than He, \ie with a
        spin--orbit splitting of the first ionic ground state
        \(n\conf{p^5\;^2P}_{j_i}\), \(j_i=\sfrac{3}{2},\sfrac{1}{2}\))
        photo-ionized with a tailored XUV pulse consisting of two
        frequencies with \(\Omega_>-\Omega_<=\omega_0\). The energy scale is that of
        the photo-electron kinetic energy, which depends on the final
        ion state, \(\conf{^2P_{\sfrac{3}{2}}}\) or
        \(\conf{^2P_{\sfrac{1}{2}}}\).  It can be seen from the
        diagram that there are four pathways to the continuum. If
        \(\omega_0\approx\Ddiff E_{\textrm{s--o}}\), two of the quantum paths
        (absorption of \(\Omega_<\) and the ion in
        \(\conf{^2P_{\sfrac{3}{2}}}\); absorption of \(\Omega_>\) and the
        ion in \(\conf{^2P_{\sfrac{1}{2}}}\)) lead to the same
        photo-electron energy.}}{
\tikzsetnextfilename{sketch}
\def\IP{3.0}
\def\DEso{1.0}
\def\harm1{5.0}
\def\photoneng{1}
\begin{tikzpicture}[scale=0.8]
  \begin{scope}
    \fill[pattern=north east lines, pattern color=dimmed] (0,0)
    rectangle ++(4,0.2) ++(0.1,-0.2) rectangle ++(4,0.2);
    \draw (0,0) node[anchor=north west]{\(\conf{^2P_{\sfrac{3}{2}}}\)} -- ++(4,0)
    ++(0.1,0) node[anchor=north west]{\(\conf{^2P_{\sfrac{1}{2}}}\)} -- ++(4,0);
    \draw[->] (-0.1,-3) node[anchor=east]{\(-I_{\mathrm{p}}\)} --
    ++(0,3) node[anchor=east]{\(0\)} --
    ++(0,5.5) node[anchor=east]{\(\varepsilon\)};
    \draw (0,-\IP) node[anchor=north west]{$\conf{^1S_0}$} -- ++(4,0)
    ++(0.1,-\DEso) node[anchor=north west]{$\conf{^1S_0}$} -- ++(4,0);
    \draw[<->] (0,-\IP) ++(4.05,0) -- ++(0,-\DEso) node[midway,anchor=east]{$\Ddiff E_{\textrm{s--o}}$};
    \draw[dimmed,domain=0:5,smooth,variable=\y,samples=100] plot ({exp(-(\y+\IP-\harm1)*(\y+\IP-\harm1)/0.1)+exp(-(\y+\IP-\harm1-\photoneng)*(\y+\IP-\harm1-\photoneng)/0.1)+1},\y);
    \draw[->,accent] (2.5,-3) -- ++(0,\harm1) node[anchor=south]{$\Omega_<$};
    \draw[->,accent] (3,-3) -- ++(0,\harm1) -- ++(0,\photoneng) node[anchor=south]{$\Omega_>$};
    \draw[dimmed,domain=0:5,smooth,variable=\y,samples=100] plot ({exp(-(\y+\IP+\DEso-\harm1)*(\y+\IP+\DEso-\harm1)/0.1)+exp(-(\y+\IP+\DEso-\harm1-\photoneng)*(\y+\IP+\DEso-\harm1-\photoneng)/0.1)+1+4.1},\y);
    \draw[->,accent] (4.1,0) ++(2.5,-4) -- ++(0,\harm1) node[anchor=south]{$\Omega_<$};
    \draw[->,accent] (4.1,0) ++(3,-4) -- ++(0,\harm1) -- ++(0,\photoneng) node[anchor=south]{$\Omega_>$};
    \draw[->,accent] (4.1,0) ++(2.5,-4) -- ++(0,\harm1) node[anchor=south]{$\Omega_<$};
    \draw[<->,accent] (4.1,0) ++(3.5,-4) ++(0,\harm1) -- ++(0,\photoneng) node[midway,anchor=west]{$\omega_0$};
  \end{scope}
\end{tikzpicture}}
  \end{floatrow}
\end{figure}
Figure \ref{fig:sketch} shows a simplified diagram of photo-ionization
of a \(n\conf{p}\) electron. The ionic ground state has a spin--orbit
splitting, which in xenon is \SI{1.3}{\electronvolt}. We ionize with a
weak XUV pulse with two frequency components, whose difference is
\(\omega_0\). Absorption of the two frequency components, \(\Omega_>\) and
\(\Omega_<\), leads to an ion in either
\(\conf{^2P^{o}_{\sfrac{3}{2}}}\) or
\(\conf{^2P^{o}_{\sfrac{1}{2}}}\), resulting in four different
pathways. If the frequency difference is equal to the spin--orbit
spacing, there will be two (indistinguishable) pathways to the same
final photo-electron energy; we call this the \emph{resonant case}. We
introduce the \emph{detuning ratio}
\(d\equiv\omega_0/\Ddiff E_{\textrm{s--o}}\), and study photoionization in the
vicinity of this resonance (\(d\approx1\)).

The calculations are performed by solving the time-dependent
Schrödinger equation (TDSE) in a limited subspace,
\begin{equation}
  \label{eqn:tdse}
  \im\partial_t\ket{\Psi(t)} = \Ham(t)\ket{\Psi(t)},
\end{equation}
where the Hamiltonian in the dipole approximation is
\begin{equation}
  \label{eqn:hamiltonian}
  \Ham(t) = \Ham_0 + \op{E}(t)z.
\end{equation}
\(\Ham_0\) is the atomic Hamiltonian, \(\op{E}(t)\) the electric
field, and \(z\) is the dipole operator for linear polarization
along the \(z\) axis. The solution is found by propagating the
initial state (the neutral ground state) to time \(t\)
\begin{equation}
  \ket{\Psi(t)} = \op{U}(t,0)\ket{\Psi_0},
\end{equation}
where the short-time propagator \(\op{U}(t+\Ddiff t,t)\) is
approximated by a \textcite{Magnus1954} propagator of fourth order
\parencite{Saad1992SJoNA,Alvermann2012}.

The time-dependent wavefunction is expanded as
\begin{equation}
  \label{eqn:expansion}
  \ket{\Psi(t)} =
  c_0(t)\ket{\Psi_0} +
  \sum_{i}\sum_{\ell}\int\diff{\varepsilon}
  c_{i}^{\ell}(t;\varepsilon)
  \ket{i \varepsilon \ell},
\end{equation}
where \(\ket{\Psi_0}\) is the ground state
\(n\conf{s^2}n\conf{p^6\;^1S_0}\) with energy \(-\Ip\), \(c_0(t)\) the
complex, time-dependent amplitude, \(i\) denotes the final state of
the ion, and \(\varepsilon \ell\) the quantum state of the photo-electron with
angular momentum \(\ell\) and energy \(\varepsilon\) (related to the momentum
\(k\) by \(\varepsilon=k^2/2\)).  The ionization channels formed by different
possible combinations of \(i\) and \(\ell\), are listed in table
\ref{tab:ion-channels}, in the case of \(jK\) coupling. \(jK\) (or
pair) coupling is defined as \parencite{Cowan1981}
\(\vec{j}_i + \Bell = \vec{K}\) and \(\vec{K} + \vec{s} = \vec{J}\),
where \(\vec{j}_i\) is the total angular momentum of the parent ion,
which couples to the angular momentum of the electron \(\Bell\) to
form an intermediate \(\vec{K}\). The levels are then written as
\(\gamma_i(^{2S+1}L_{j_i})k\ell\;^{2S+1}[K]_J\), where \(\gamma_i\) is the electron
configuration of the ion.

The \emph{Ansatz} \eqref{eqn:expansion} turns the
TDSE~\eqref{eqn:tdse} into a set of coupled ordinary differential
equations (ODE):
\begin{equation}
  \label{eqn:tdse-expansion}
  \im\partial_t\vec{c}(t) = \mat{H}(t)\vec{c}(t),
  \tag{\ref{eqn:tdse}*}
\end{equation}
where the vector \(\vec{c}(t)\) consists of the expansion coefficients
in \eqref{eqn:expansion}, and the \emph{Hamiltonian matrix} is given
by
\begin{equation}
  \begin{aligned}
    \mat{H}(t) =&
    -\ketbra[\Ip]{\Psi_0}{\Psi_0}
    + \sum_{i\ell}\int\diff{\varepsilon}
    \ketbra[\varepsilon]{i\varepsilon\ell}{i\varepsilon\ell}\\
    &+
    \op{E}(t)\left[
      \sum_{i\ell}\int\diff{\varepsilon}
      \ket{i\varepsilon\ell}\matrixel{i\varepsilon\ell}{z}{\Psi_0}\bra{\Psi_0}
      +
      \sum_{i'\ell'}\int\diff{\varepsilon'}
      \ket{i\varepsilon\ell}\matrixel{i\varepsilon\ell}{z}{i'\varepsilon'\ell'}\bra{i'\varepsilon'\ell'}
      +\cc
    \right].
  \end{aligned}
  \label{eqn:hamiltonian-expansion}
  \tag{\ref{eqn:hamiltonian}*}
\end{equation}
In the field-free basis, \(\mat{H}_0\) is simply a diagonal matrix,
with the energies of the photo-electron with respect to the lowest
ionization threshold as matrix elements. The interaction term couples
the ground state to the continua and the continua to each other. In
the weak-field limit, however, the partial-wave expansion is
restricted to total angular momentum \(J\leq1\), \ie no multi-photon
processes are considered. Furthermore, ionization is only allowed from
the outer \(n\conf{p}\) shell (photo-electron energies in the range
\SIrange{0}{11}{eV} in the case of xenon), to avoid autoionization of
embedded Rydberg states in the vicinity of the
\(n\conf{s}n\conf{p^6\;^2S_{\sfrac{1}{2}}}\) threshold (that is,
channels 6 and 7 in table~\ref{tab:ion-channels} need not be
considered). We also neglect mixing of singlet and triplet
terms. Thus, the only non-zero matrix elements of the dipole operator
\(z\) are \(\matrixel{i \varepsilon \ell}{z}{\Psi_0}\) (and the complex
conjugate). The basis functions (\(\ket{\Psi_0}\) and
\(\ket{i\varepsilon\ell}\)) and the dipole matrix elements
(\(\matrixel{i \varepsilon \ell}{z}{\Psi_0}\)) are determined using \program{ATSP2K}
(multi-configurational Hartree--Fock; \cite{FroeseFischer2007CPC}) and
\program{BSR} (close-coupling \(R\)-matrix approach;
\cite{Zatsarinny2006,Zatsarinny2009}). The dipole matrix elements are
spin-averaged by \program{BSR}.

The analysis of the coherence is made using the density matrix
formalism [\cite[§14]{Landau1977quant}], where the full \emph{density
  matrix} operator is formed from the wavefunction \(\ket{\Psi(t)}\)
obtained by solving \eqref{eqn:tdse} (time dependence \(t\) omitted
below, for brevity)
\begin{equation}
  \label{eqn:density-matrix-operator}
  \rho_T = \ketbra{\Psi}{\Psi},
\end{equation}
with matrix elements of the continuum block
\begin{equation}
  \label{eqn:continuum-densities}
  \rho_{i_1i_2}^{\ell_1\ell_2}(\varepsilon_1,\varepsilon_2) \equiv
  c_{i_1}^{\ell_1}(\varepsilon_1)
  c_{i_2}^{\ell_2*}(\varepsilon_2).
\end{equation}
We reduce this density matrix\index{reduced density matrix} to an
ion--channel density matrix by first taking the trace over the
photo-electron energy \(\varepsilon\):
\begin{equation}
  \label{eqn:ion-channel-density-matrix}
  \begin{aligned}
    \rho_{i_1i_2}^{\ell_1\ell_2}&\equiv
    \int\diff{\varepsilon}
    \braket{\varepsilon}{\Psi}\braket{\Psi}{\varepsilon}
    =\int\diff{\varepsilon}
    \sum_{i_1i_2}\sum_{\ell_1\ell_2}
    \int\diff{\varepsilon_1}\diff{\varepsilon_2}
    \braket{\varepsilon}{i_1\varepsilon_1\ell_1}
    c_{i_1}^{\ell_1}(\varepsilon_1)c_{i_2}^{\ell_2*}(\varepsilon_2)
    \braket{i_2\varepsilon_2\ell_2}{\varepsilon}\\
    &=
    \sum_{i_1i_2}\sum_{\ell_1\ell_2}
    \int\diff{\varepsilon}
    \ket{i_1\ell_1}
    \rho_{i_1i_2}^{\ell_1\ell_2}(\varepsilon,\varepsilon)
    \bra{i_2\ell_2}.
  \end{aligned}
\end{equation}
Finally, we construct the ion density matrix by tracing over the
photo-electron angular momenta:
\begin{equation}
  \label{eqn:ion-density-matrix}
  \begin{aligned}
    \rho_{i_1i_2}&\equiv
    \sum_\ell
    \sum_{i_1i_2}\sum_{\ell_1\ell_2}
    \braket{\ell}{i_1\ell_1}
    \rho_{i_1i_2}^{\ell_1\ell_2}
    \braket{i_2\ell_2}{\ell}
    =
    \sum_\ell
    \sum_{i_1i_2}
    \ket{i_1}
    \rho_{i_1i_2}^{\ell\ell}
    \bra{i_2}.
  \end{aligned}
\end{equation}
The diagonal elements (\(\rho_{mm}\)) of this matrix are the populations
in each of the ionic states, while the off-diagonal elements
(\(\rho_{mn}\)) contain the coherences between the ionic states. The only
non-zero off-diagonal elements are those corresponding to channels for
which all quantum numbers are the same (except for the angular
momentum of the ion); \ie only \(\rho_{35} = \conj{\rho}_{53}\neq0\).
Decoherence due to decay (through dipole-forbidden interaction) from
the upper ionic state to the lower, is neglected.
\section{The bichromatic case}
\label{sec:bichromatic}
\begin{figure}[tb]
  \centering
  \includegraphics{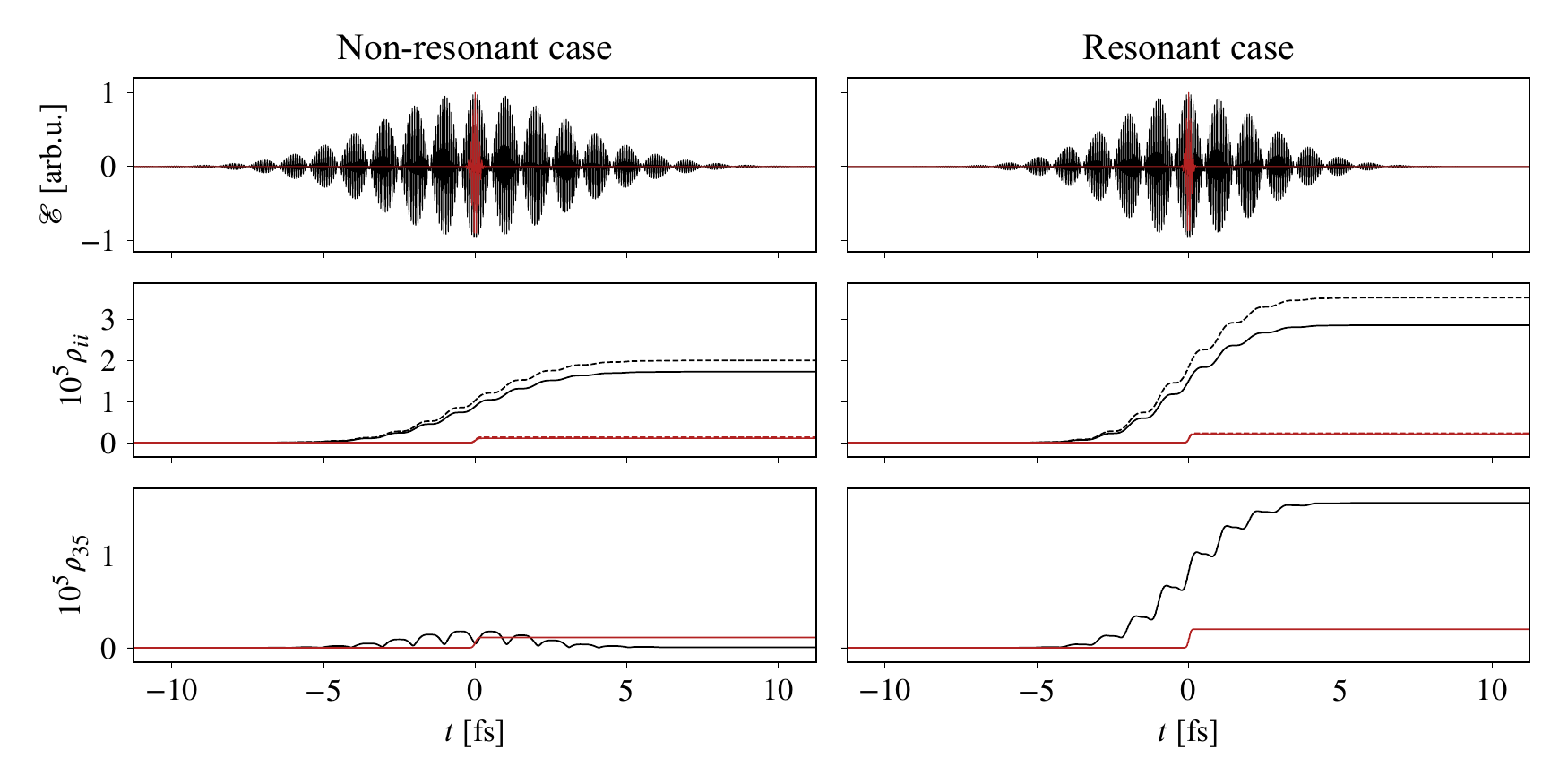}
  \caption{\label{fig:real-time-coherence} Real-time coherence
    build-up, for the case of ionization with two harmonics, 13 and 14
    of a fundamental frequency \(d\Ddiff E_{\textrm{s--o}}\), in the
    non-resonant case (\(d=1.3\)) left and the resonant case (\(d=1\))
    right. The upper panels show the driving fields used; the red
    curve corresponds to a pulse duration (FWHM of the temporal
    intensity profile) of \SI{500}{\atto\second} while the black curve
    corresponds to a pulse duration of \SI{15}{\femto\second}. The
    middle panels show the populations in the residual ionic substates
    (solid: contribution from channel 3 to
    \(\conf{^2P^o_{\sfrac{3}{2}}}\); dashed: contribution from channel
    5 to \(\conf{^2P^o_{\sfrac{1}{2}}}\)), which increase with
    time. The lower panels show the induced coherence between the
    ionic substates, which is built up over time. For the short-pulse
    case, there is always coherence left at the end of the pulse,
    while for the longer pulse duration, the resonance criterion has
    to be fulfilled \(d\approx1\) for this to happen. The lower population
    in the non-resonant case is explained by the decrease in
    photoionization cross-section with increasing photon energy.}
\end{figure}
We investigate the real-time build-up of coherence with a XUV pulse of
short or long duration in the non-resonant case
(figure~\ref{fig:real-time-coherence}, left) and the resonant case
(figure~\ref{fig:real-time-coherence}, right). The electromagnetic
fields are presented in the upper panels. In both cases, they consist
of harmonics 13 \& 14 of a fundamental driving field, and a peak
intensity of \SI{e8}{\watt\per\centi\meter\squared}. The short-pulse
duration is \SI{500}{\atto\second}, while the long-pulse duration is
\SI{15}{\femto\second} resulting in the formation of a periodic
beating of the XUV pulse. Regardless of the pulse duration, the
population in the ionic substates (middle shown in solid and dashed
lines) increases as the pulse ionizes the atom. The population is
proportional to the integral of the pulse intensity, hence the
appearance of steps in the population.

\subsection{The non-resonant case.}
\label{sec:bichromatic-non-resonant}
We first consider the non-resonant case (left panels of figure
\ref{fig:real-time-coherence}), where the fundamental driving
frequency is \(\omega_0=1.3\Ddiff E_{\textrm{s--o}}\). For a short pulse
duration, the coherence increases during the interaction and stays
constant after the pulse has passed. In contrast, for long pulse
duration, the coherence first builds up, and then vanishes at the end
of the pulse. The decoherence time, \ie the time from the onset of the
pulse to the decrease of the coherence (see lower panel of figure
\ref{fig:real-time-coherence}), is approximately equal to the quantum
beat period \(T=2\cpi/\Ddiff E_{\textrm{s--o}}\) of the ionic
substates; for xenon with a spin--orbit splitting of
\SI{1.3}{\electronvolt}, it is \SI{3.2}{\femto\second}. This is
similar to what was observed by \textcite{Goulielmakis2010}, where the
interference pattern, present in the transient absorption signal,
disappeared when the ionizing infrared pulse exceeded the quantum beat
period of \SI{6.2}{\femto\second} of krypton.

\subsection{The resonant case. A generalized quantum beat condition.}
\label{sec:orgd61bd89}
In the resonant case (right panels of figure
\ref{fig:real-time-coherence}), the situation is completely
different. The coherence is built up during all the interaction time,
and remains after the end of the pulse. We can understand the
existence of this resonance condition, using a simple
two-excited-state model. The two excited states (labelled \(\ket{1}\)
and \(\ket{2}\), for simplicity) correspond to the ionic substates,
the populations of which result from a periodic sequence of ionization
events (occurring at times \(t_k\), \(k\in\Natural\)) from the ground
state (labelled \(\ket{0}\)). We express this model using an
inhomogeneous TDSE for the excited superposition
\(\ket{\Psi(t)}=c_1(t)\ket{1}+c_2(t)\ket{2}\), where the ground state
constitutes a source term:
\begin{equation}
  \im\partial_t\ket{\Psi(t)} = \mat{H}_0\ket{\Psi(t)} +
  \mathcal{E}(t)(z_{10}\ket{1}+z_{20}\ket{2}),
  \quad
  \mat{H}_0\equiv
  E_1\ketbra{1}{1}+E_2\ketbra{2}{2},
  \quad
  z_{i0}\equiv\matrixel{i}{z}{0}.
  \label{eqn:two-level-system}
\end{equation}
Assuming no initial excited population, the solution of
\eqref{eqn:two-level-system} is given by
\begin{equation}
  \begin{aligned}
    \ket{\Psi(t)}
    &=
    \exp(\im\mat{H}_0t)
    \left\{
      \int^t\diff{t'}
      \op{E}(t')
      [\ce^{-\im E_1t'}z_{10}\ket{1}
      +\ce^{-\im E_2t'}z_{20}\ket{2}]
    \right\}
  \end{aligned}
  \label{eqn:two-level-solution}
\end{equation}
If we assume that \(\op{E}(t)\) is a train of pulses, separated by
\(\Ddiff t\equiv t_k - t_{k-1}\), we can write the field as
\[
  \op{E}(t) =
  \tilde{\op{E}}(t)\left[
    \sum_k\delta(t_k)\ce^{\im\phi_{\textrm{XUV}}(t_k)}
  \right],
\]
where \(\tilde{\op{E}}(t)\) is a slowly varying envelope. The solution
to the two-excited-state system in this case is
\begin{equation}
  \begin{aligned}
    \ket{\Psi(t)}
    &=
    \exp(\im\mat{H}_0t)
    \sum_k
    \tilde{\op{E}}(t_k)
    \ce^{\im\phi_{\textrm{XUV}}(t_k)}
    \left[
      \ce^{-\im E_1t_k}z_{10}\ket{1}
      +\ce^{-\im E_2t_k}z_{20}\ket{2}
    \right]\\
    &=
    \sum_k
    \tilde{\op{E}}(t_k)
    \ce^{\im\phi_{\textrm{XUV}}(t_k)}
    \left[
      \ce^{-\im E_1(t_k-t)}z_{10}\ket{1}
      +\ce^{-\im E_2(t_k-t)}z_{20}\ket{2}
    \right]\\
    &=
    \sum_k
    \tilde{\op{E}}(t_k)
    \ce^{\im[\phi_{\textrm{XUV}}(t_k) - E_1(t_k-t)]}z_{10}
    \left[
      \ket{1}
      +\tilde{z}\ce^{-\im \Ddiff E (t_k-t)}\ket{2}
    \right],
  \end{aligned}
  \label{eqn:two-level-solution-train}
  \tag{\ref{eqn:two-level-solution}*}
\end{equation}
where \(\tilde{z}\equiv\matrixel{2}{z}{0}/\matrixel{1}{z}{0}\) is the
relative dipole matrix element independent of the instant of
ionization, and --- assuming ionization solely into an unstructured
continuum --- independent of final electron energy and
\(\Ddiff E \equiv E_2 - E_1\), as well.

For the two substates to remain coherent, we require that no dephasing
is introduced by pulses in the train. Since subsequent pulses are
separated in time by \(\Ddiff t\), this is equivalent to requiring
that the phase argument in \eqref{eqn:two-level-solution-train}
fulfills \(\Ddiff E(t_k-t_a)=\Ddiff E\Ddiff t(k-a)=2\cpi\), \(k,a\in\Natural\), or
\begin{equation}
  \label{eqn:quantum-beat-condition}
  \Ddiff t = \frac{2\cpi r}{\Ddiff E}=rT,
\end{equation}
is fulfilled, for any integer \(r\). \(\Ddiff t\) is a multiple of the
quantum beat period for an energy separation \(\Ddiff E\), which does
not depend on the duration of the XUV pulse. In the spectral domain,
this corresponds to requiring the final electron kinetic energy to be
the same.

\subsection{Maximum coherence achievable. Degree of coherence.}
\label{sec:maximum-coherence}
\begin{figure}[tb]
  \centering
  \includegraphics{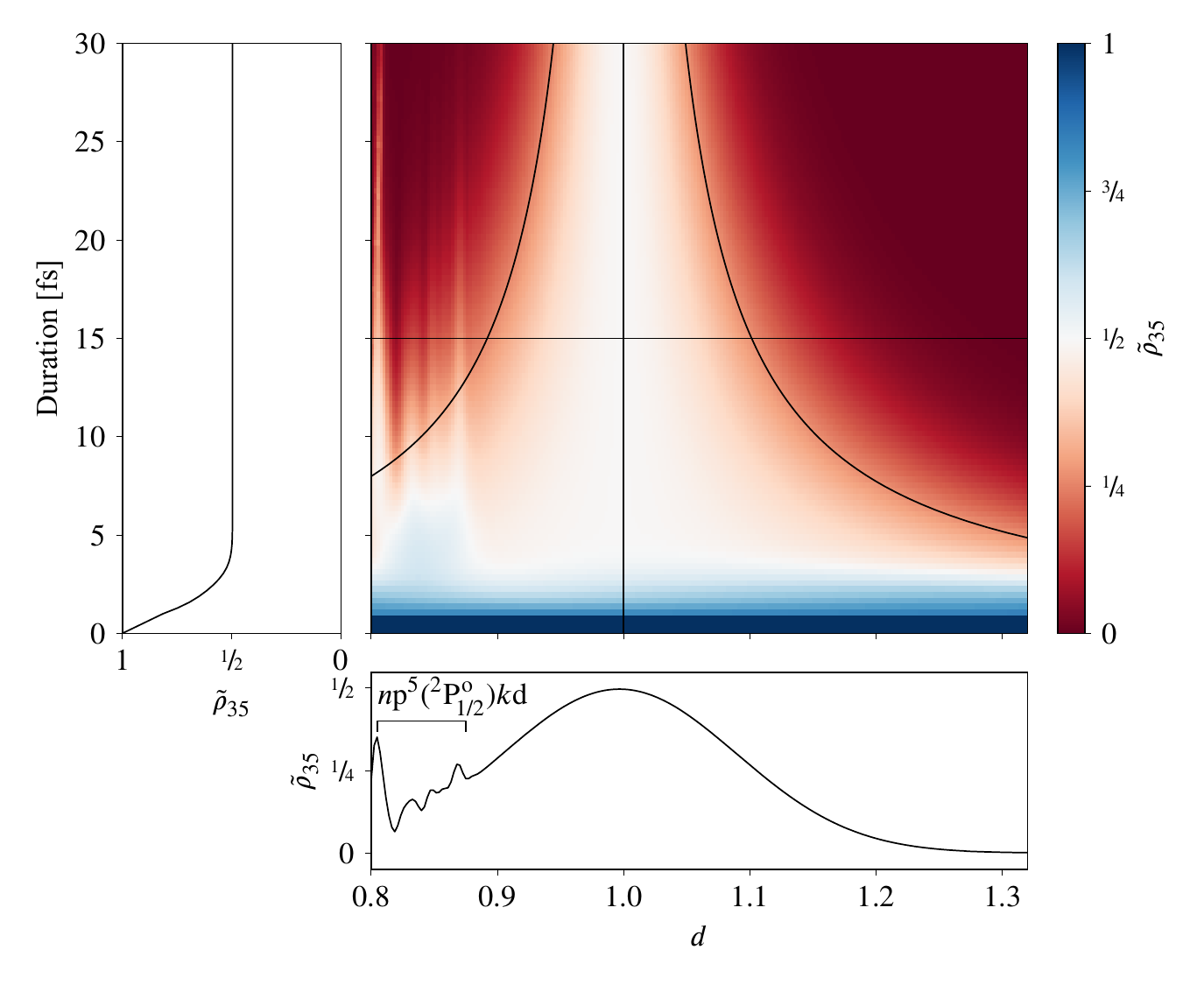}
  \caption{\label{fig:duration-detuning-map} Degree of coherence as a
    function of pulse duration (FWHM of temporal intensity profile)
    and detuning ratio \(d\). The hyperbolas in the main panel
    indicate the `coherence bandwidth' within which
    \(\tilde{\rho}_{35}>(2\sqrt{\ce})^{-1}\). The black
    horizontal/vertical lines mark lineouts at constant
    duration/detuning ratio, which are shown in the lower/left panels.
    As the pulse duration exceeds the quantum beat period, the peaks
    become spectrally resolvable, and the total overlap at the central
    frequency becomes half of that at vanishing pulse duration. The
    peaks appearing below \(d\approx0.9\) occur when the lower harmonic
    excites autoionizing Rydberg states between
    \(\conf{^2P_{\sfrac{3}{2}}}\) and \(\conf{^2P_{\sfrac{1}{2}}}\)
    thresholds.}
\end{figure}
As seen in figure~\ref{fig:real-time-coherence}, given that the
resonance condition \(d=1\) is fulfilled, the coherence between the
ionization channels seem to increase as long as the pulse is of
appreciable amplitude. What is the maximum coherence achievable using
this scheme? The theoretical maximum coherence
\(\abs{\rho_{mn}}=\abs{c_m\conj{c_n}}\), is bounded by the
Cauchy--Schwartz inequality:
\begin{equation}
  \label{eqn:cauchy--schwartz}
  \abs{\rho_{mn}}^2 \leq \rho_{mm}^2\rho_{nn}^2,
\end{equation}
where \(\rho_{ii}\) is the probability of finding the system in state
\(\ket{i}\), which obviously cannot exceed unity. From
\eqref{eqn:cauchy--schwartz}, it is natural to introduce the
\emph{degree of coherence}:
\begin{equation}
  \label{eqn:degree-of-coherence}
  \tilde{\rho}_{mn} \equiv \frac{\abs{\rho_{mn}}}{\sqrt{{\rho_{mm}}{\rho_{nn}}}},
\end{equation}
which normalizes the coherence between two ions to their respective
populations. This quantity is useful since, even though the
populations in two states are minuscule (as is the case in
figure~\ref{fig:real-time-coherence}) and thereby also the absolute
coherence, they may be fully coherent \emph{with respect to each
  other}. If this is the case, the degree of coherence will be
unity.

Figure~\ref{fig:duration-detuning-map} shows the degree of coherence,
as a function of XUV pulse duration and detuning ratio. For short
pulse durations, this quantity is larger than \sfrac{1}{2} (left panel
of figure \ref{fig:duration-detuning-map}). In this regime, the
interaction with the XUV pulse occurs within one quantum beat period
(\SI{3.2}{\femto\second}), and the four pathways into the continuum
(indicated in figure \ref{fig:sketch}) have a partial spectral
overlap. For larger pulse durations, two of the pathways become
distinguishable, and do not contribute to the coherence between the
ionic substates. The two remaining pathways, namely via \(\Omega_<\)
leaving the ion in \(\conf{^2P^{o}_{\sfrac{3}{2}}}\), and via
\(\Omega_>\) leaving the ion in \(\conf{^2P^{o}_{\sfrac{1}{2}}}\), cannot
be distinguished when measuring the photo-electron. Provided the
resonance condition \(d=1\) is met, the maximum degree of coherence is
\sfrac{1}{2}. Complete decoherence always occurs in the long-pulse
limit if \(d\neq1\) since the quantum pathways are distinguishable. As we
will see below, the maximum degree of coherence can be increased by
adding more colours to the ionizing field.

\section{The multi-colour case --- Ionization by an attosecond pulse
  train}
\label{sec:multi-colour}
We consider now the effect of ionization with an attosecond pulse
train, by including additional harmonic components. To focus on this
aspect of the problem, we use a simplified model, where the dipole
matrix elements for ionization are replaced with Heaviside functions
(this is an approximation of a flat continuum, \ie no resonances
present):
\begin{equation}
  z(\varepsilon) = \theta(\varepsilon),
\end{equation}
where \(\varepsilon\), as before, is the energy of the continuum electron. We
still use the \emph{Ansatz} \eqref{eqn:expansion}, albeit with a
compact notation only considering different channels \(n\):
\begin{equation}
  \label{eqn:expansion-simplified}
  \ket{\Psi(t)}
  =
  c_0(t)\ket{\Psi_0}
  +\sum_n \int \diff{\varepsilon}
  c_n(t;\varepsilon)\ket{n\varepsilon}.
\end{equation}
Inserting this in the Schrödinger equation and applying first-order
time-dependent perturbation theory (\ie the ground state is unaffected
by the weak-field ionization; \(c_0(t)=1\)), the solution reads
\begin{equation}
  \label{eqn:spectral-model-solution}
  c_n(t)
  =-\im\theta(\varepsilon)
  \int_{-\infty}^t\diff{t'}
  \mathcal{E}(t')
  \exp(\im E_n t').
\end{equation}
Evaluating at the time of measurement (\(t=+\infty\)), we see that the
coefficient becomes the Fourier transform of the driving field,
evaluated at \(-E_n - \varepsilon\). From this, we get the coherence between two
channels \((m,n)\) as
\begin{equation}
  \label{eqn:simplified-coherence}
  \rho_{mn}
  =-\int\diff{\varepsilon}\theta(\varepsilon)
  \conj{\hat{\mathcal{E}}}(E_m+\varepsilon)
  \hat{\mathcal{E}}(-E_n-\varepsilon),
\end{equation}
where \(\hat{\mathcal{E}}(\omega)\) designates the Fourier transform of
\(\mathcal{E}(t)\). This expression shows that the coherence primarily
arises from a correlation of the field with itself shifted by the
energy difference \(\Ddiff E \equiv E_m-E_n\).

\begin{figure}[htb]
  \centering
  \includegraphics{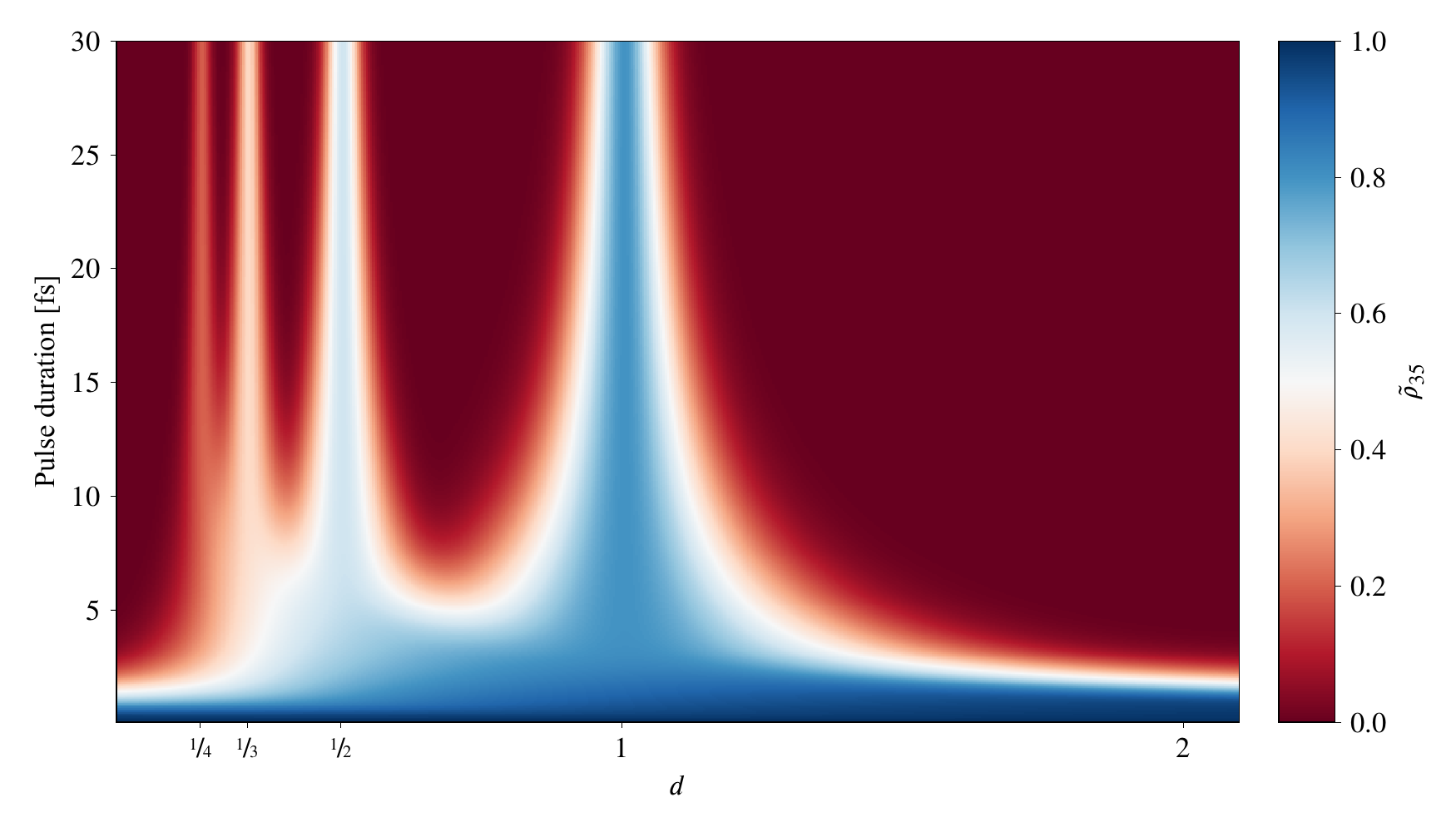}
  \caption{The effect of using five harmonics: At \(d=1\), 4 harmonics
    out of 5 will yield photo-electron peaks that overlap with those
    shifted by the spin--orbit splitting. The peaks appearing at
    \(d=\sfrac{1}{4},\sfrac{1}{3},\sfrac{1}{2}\) correspond to 1, 2,
    and 3 harmonics resulting in overlapping photo-electron peaks,
    respectively. In the time domain, this corresponds to ionizing
    pulses arriving at every \(r=4,3,2\) quantum beat periods.}
  \label{fig:multiple-harms}
\end{figure}
In the long-pulse limit, the spectral components of
\(\hat{\op{E}}(\omega)\) will become very narrow. If the ionizing field is
an attosecond pulse train, consisting of the \(n_q = q_2-q_1+1\)
successive harmonic orders \(q\in\{q_1..q_2\}\), the shift has to be
precisely an integer amount of photons (\(\Delta E=N\omega_0\)) for the
integrand of \eqref{eqn:simplified-coherence} to be non-zero. Assuming
\(q_1\omega_0, q_2\omega_0>E_m,E_n\), \ie all constituent harmonic orders reach
above both ionization thresholds, we have in total \(2n_q\) pathways
into the continuum. In the case of an integer photon shift,
\(2(n_q-N)\) of these pathways will overlap, which means
\begin{equation}
  \label{eqn:multi-colour-coherence}
  \tilde{\rho}_{mn}=\frac{n_q-N}{n_q}.
\end{equation}
From \eqref{eqn:multi-colour-coherence}, we see that by adding more
and more colours, we can increase the degree of coherence towards
unity. This is illustrated in figure \ref{fig:multiple-harms} for the
case of five harmonics; the maximum degree of coherence is indeed
\sfrac{4}{5}, which occurs at \(d=1\).

Figure~\ref{fig:multiple-harms} also serves the purpose of
illustrating the generalization of the quantum beat condition
\eqref{eqn:quantum-beat-condition}; \eg the case \(r=2\) corresponds
in the time domain to ionizing pulses arriving every other quantum
beat period. Such a field can be realized through red--blue HHG
(resulting in odd \emph{and} even harmonic orders) from a fundamental
frequency \(\Ddiff E_{\textrm{s--o}}/2\) (for Xe with
\(\Ddiff E_{\textrm{s--o}}\approx\SI{1.3}{\electronvolt}\), a fundamental
driving wavelength \(\lambda\approx\SI{1.9}{\micro\meter}\) would thus be
necessary). From this we see that in the spectral domain, the
generalized quantum beat condition \eqref{eqn:quantum-beat-condition}
is simply \(d=\sfrac{1}{r}\), \(r\in\Integer\).

Finally, we have also checked that in the long-pulse limit, the case
of \(d=\sfrac{1}{2}\), using only odd-order harmonics is no different
from the case of \(d=1\), using even- and odd-order harmonics. Thus
the change of parity (phase difference) of consecutive pulses in the
attosecond pulse train has no impact on the coherence, and we conclude
that the degree of coherence is essentially dictated by the amount of
indiscernible pathways.
\section{Conclusion}
\label{sec:org67cfded}
In summary, we have shown that it is possible to induce coherence
between two ionic substates using pulses of duration longer than the
quantum beat time of their superposition. This is possible, provided a
resonance condition is fulfilled, namely that the driving field has at
least two frequency components spaced by precisely the energy
difference of the levels of interest. This result shows that when the
electron wave packets arising from different pathways have the same
kinetic energy, we cannot know which way the ionization occurred. This
situation is reminiscent of a Young's double slit experiment [see for
instance \textcite{Arndt2005}], with the two harmonic orders playing
the roles of the two slits. It has to be noted, however, that no
interference can be detected in the photo-electron signal, unless the
ions are brought to the same final state via some mechanism [\eg
\textcite{Goulielmakis2010} did this by further exciting from the fine
structure superposition using an XUV pulse]. Otherwise, it would be
possible to detect the ions and the photo-electrons in coincidence
mode and establishing the ionization pathway.
\section{Acknowledgments}
\label{sec:acknowledgements}
We acknowledge the help of Oleg Zatsarinny, and would like to thank
Andreas Buchleitner, Andreas Wacker, Kevin Dunseath, Mikhail Ivanov,
Tobias Brünner, and Tomas Brage for fruitful discussions. This work
was supported by the Swedish Foundation for Strategic Research, the
Swedish Research Council, the Knut and Alice Wallenberg Foundation,
the European Research Council (PALP), and by funding from the
NSF under grant PHY-1307083.
\printbibliography
\end{document}